\documentclass[aps,amsmath,amssymb,reprint]{revtex4-1}
\usepackage{graphics}
\usepackage{graphicx}      
\usepackage{epsfig}
\usepackage{amsmath}
\usepackage{array}
\usepackage{amssymb}
\usepackage{dcolumn}       
\usepackage{bm}            
\usepackage{array}
\usepackage{booktabs}
\usepackage{ulem}
\usepackage{physics}     
\usepackage{subcaption}
\usepackage[utf8]{inputenc}
\usepackage{url}

\usepackage[linktocpage=true,
    colorlinks=true,
    pdfborder={0 0 0},
    linkcolor=blue,
    citecolor=red,
    filecolor=yellow,
    urlcolor=blue,
    bookmarks,
    pdfauthor={},
]{hyperref}

\usepackage[dvipsnames]{xcolor}

\newcommand{\affil}{Laboratory for Materials and Structures, Institute of Innovative Research, \\ Tokyo Institute of Technology, Tokyo, 152-8550, Japan.\\
Quemix Inc., Tokyo, 103-0027, Japan.}

\begin{document}

\title{Implementation of quantum imaginary-time evolution method on NISQ devices: Nonlocal approximation}

\author{Hirofumi Nishi} \affiliation{\affil}  
\author{Taichi Kosugi} \affiliation{\affil}
\author{Yu-ichiro Matsushita} \affiliation{\affil} 

\date{\today}

\begin{abstract}
The imaginary-time evolution method is widely known to be efficient for obtaining the ground state in quantum many-body problems on a classical computer. A recently proposed quantum imaginary-time evolution method (QITE) faces problems of deep circuit depth and difficulty in the implementation on noisy intermediate-scale quantum (NISQ) devices. In this study, a nonlocal approximation is developed to tackle this difficulty. We found that by removing the locality condition or local approximation (LA), which was imposed when the imaginary-time evolution operator is converted to a unitary operator, the quantum circuit depth is significantly reduced. We propose two-step approximation methods based on a nonlocality condition: extended LA (eLA) and nonlocal approximation (NLA). To confirm the validity of eLA and NLA, we apply them to the max-cut problem of an unweighted 3-regular graph and a weighted fully connected graph; we comparatively evaluate the performances of LA, eLA, and NLA. The eLA and NLA methods require far fewer circuit depths than LA to maintain the same level of computational accuracy. Further, we developed a ``compression'' method of the quantum circuit for the imaginary-time steps as a method to further reduce the circuit depth in the QITE method. The eLA, NLA, and the compression method introduced in this study allow us to reduce the circuit depth and the accumulation of error caused by the gate operation significantly and pave the way for implementing the QITE method on NISQ devices.
\end{abstract}

\maketitle

Quantum computers, initially proposed by Feynmann\cite{feynman1999simulating}, were unveiled by Deutsch\cite{deutsch1992rapid}, Grover\cite{grover1996fast}, and Shor\cite{shor1999polynomial} to have great potential that could overwhelmingly surpasses classical computers. In addition, the news of Google's demonstration of quantum supremacy in 2019 has spread around the world\cite{arute2019quantum} and expectations for the realization of practical quantum computers are increasing. One of the most promising problems for quantum computers is combinatorial optimization, which is a NP-hard problem\cite{hyafil1973graph}. Combinatorial optimization problems are closely related to our daily lives, and they include the traveling salesman problem\cite{lawler1985traveling}, scheduling problem\cite{ho2004genace}, SAT (satisfiability problem) solver\cite{karp1972reducibility}, among others. For these combinatorial optimization problems, Grover's algorithm is already known to improve the computational cost with quadratic speedup compared to classical computers\cite{durr1996quantum,baritompa2005grover}.

Under these circumstances, it is challenging for researchers all over the world to employ existing or near-future quantum computers to achieve tasks that are very difficult or impossible using classical computers. Currently available quantum computers are noisy intermediate-scale quantum (NISQ) devices\cite{preskill2018quantum}. Further, conventional quantum algorithms require many gate operations, such as Grover's algorithm, and they cannot be implemented on NISQ devices with no error correction and short coherence time. Recently, classical-quantum hybrid algorithms called variational quantum eigensolver (VQE)\cite{peruzzo2014variational,mcclean2016theory}, and quantum approximate optimization algorithm (QAOA)\cite{farhi2014quantum,otterbach2017unsupervised,moll2018quantum,wang2018quantum, Hadfield2019FromTQ, guerreschi2019qaoa} have been proposed for NISQ devices. In these methods, ansatz states with parameters are implemented as quantum circuits, and the parameters included in the ansatz states are optimized on a classical computer. While VQE and QAOA can be realized with a limited number of quantum operations and have good noise tolerance, it is difficult to determine the ansatz states properly and converge high-dimensional parameters\cite{mcclean2018barren}.

For quantum many-body problems, an imaginary-time evolution method is a known computational method to identify the ground state. The imaginary-time evolution method selectively extracts the ground state component by performing time evolution in the direction of imaginary time. Various combinatorial optimization problems are converted to a Hamiltonian format, and their corresponding Hamiltonian is derived\cite{Lucas2014IsingFO}. Thus, it is possible to solve the combinatorial optimization problem using the imaginary-time evolution method. 

The implementation of the imaginary-time evolution method on a quantum computer involves a critical problem in that the imaginary-time evolution operator is a nonunitary operator, and therefore, it cannot implement the imaginary-time evolution method on a quantum computer in its current state. To overcome this challenge, two imaginary-time evolution methods--- one that assumes an ansatz state and another that does not---were proposed in this study. The method that assumes the ansatz state traces the imaginary-time evolution of the parameters contained in the ansatz state\cite{mcardle2019variational,stokes2019quantum,david2020avoiding}. The other method introduces a unitary operation to reproduce the state on which the imaginary-time evolution operator has acted accurately\cite{motta2020determining,yeter2019practical,beach2019making}. The latter quantum imaginary-time evolution (QITE) method is considered an efficient approach to the optimization problem because it does not need to assume an ansatz state; further, there is no problem of convergence of high-dimensional parameters, even when compared to QAOA.

We focus on the QITE method without the ansatz assumption and apply it to the optimization problems. The QITE method proposed in the previous research has problems with circuit depth and computational cost; even a simple one-dimensional Ising model requires $4^4 = 256$ fourth-order tensor-product operators\cite{motta2020determining}. Further, more complex problems are challenging to implement on NISQ devices.

Therefore, we propose two approximations and one computational technique to overcome this difficulty. We succeeded in significantly reducing the quantum circuit depth of the QITE method, and we applied the developed algorithms to the max-cut problem, which is an NP-hard problem. For the max-cut problem, we chose an unweighted 3-regular graph and a weighted fully connected graph. The latter is a problem known as the classification problem in the context of unsupervised machine learning\cite{jain1999data,jain1988algorithms}.

\section*{Method}
\subsection*{Unitarization of imaginary-time evolution operators}
\begin{figure*}[tb]
   \begin{center}
       \includegraphics[bb=0 0 882 531,scale=0.55]{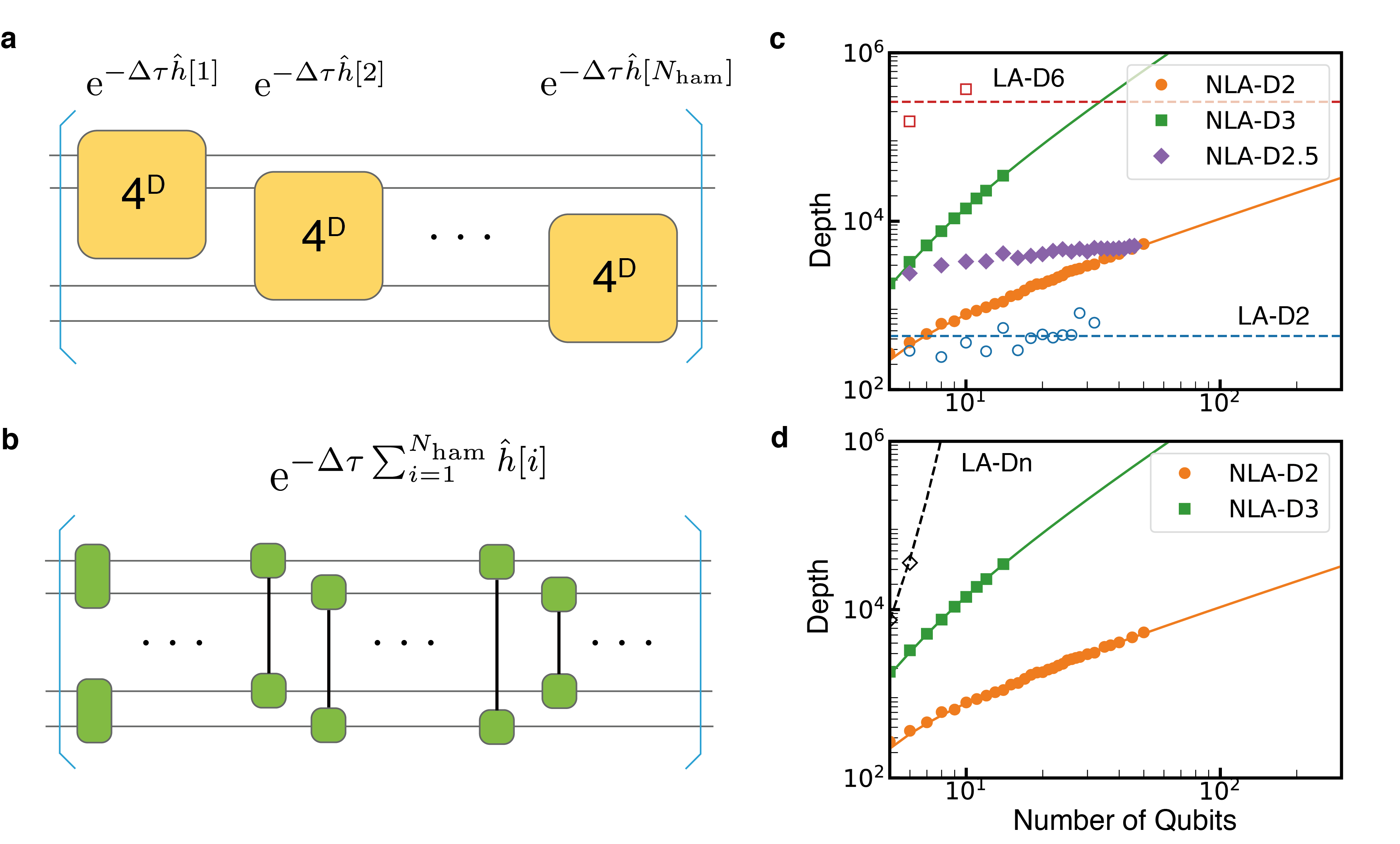}
       \caption{(a) Quantum circuit diagram for one imaginary-time step of LA. The horizontal line represents each qubit, and the yellow box represents $4^D$ gating operations on the straddling qubits. (b) Quantum circuit diagram for one imaginary-time step of the NLA (domain size $D = 2$). The green boxes and vertical lines connecting them represent a second-order tensor product operation on the two straddling qubits, with one imaginary-time step containing ${}_{N_{{\rm bit}}} C _{D}$ of second-order tensor products. The dependence of the quantum circuit depth for one imaginary-time step of the max-cut problem in the 3-regular graph (c) and the fully connected graph (d) as a function of the number of qubits.
       \label{img:circ_depth}}
   \end{center}
\end{figure*}

Consider a scenario wherein a Hamiltonian $\hat{H}$ is given for the optimization problem considered in this study. The Hamiltonian $\hat{H}$ is expressed as the summation of some partial Hamiltonians $\hat{h}[m]$ as $\hat{H}=\sum_{m = 1}^{N_{{\rm ham}}}\hat{h}[m]$, where $N_{\rm ham}$ is the number of the partial Hamiltonians.
The max-cut problem, which is a computational target of this work, is represented by the Hamiltonian in the form of Ising spins and can be mapped to the Pauli operator representation for qubits in a straightforward manner. 
In the case of the Hamiltonian of quantum chemistry, each partial Hamiltonian can be mapped to the Pauli-operator representation on qubits via the Bravyi--Kitaev representation \cite{bravyi2002fermionic} or Jordan--Wigner representation \cite{jordan1928uber}. 

For a given Hamiltonian, the ground state is obtained by using the imaginary-time evolution method. We apply the imaginary-time evolution operator defined by $\mathrm{e}^{-{\tau\hat{H}}}$, where $\tau$ is the imaginary time to reach the initial ($\tau = 0$) state of the system, $|\Psi(\tau = 0)\rangle$; and  $\mathrm{e}^{-\tau\hat{H}}|\Psi(\tau = 0)\rangle$. 
The imaginary-time evolution operator is decomposed by a first-order Suzuki--Trotter decomposition into ones with a small imaginary-time step $\Delta\tau$ ($\tau\equiv\Delta\tau\times N_{\rm step}$) of the individual partial Hamiltonians $\hat{h}[m]$.
\begin{eqnarray*}
\mathrm{e}^{-\tau\hat{H}} = \prod_{n = 1}^{N_{\rm step}} \prod_{m = 1}^{N_{\rm ham}} \mathrm{e}^{-\Delta \tau \hat{h}[m]}
+\mathcal{O}(\Delta\tau^2) \,\, .
\end{eqnarray*}
Because the operators of the imaginary-time evolution are nonunitary, they cannot be directly implemented as a gate operation on a quantum computer.
In the QITE method, the unitary operator $e^{-i\Delta \tau \hat{A}_n[m]}$
is defined such that it reproduces the state $\mathrm{e}^{-\Delta \tau\hat{H}}|{\Psi_n}\rangle$ for a given state $|{\Psi_n}\rangle\equiv |{\Psi(\tau=n\Delta\tau)}\rangle$. We determine the Hermitian operator $\hat{A}_n[m]$ that minimizes the following residual norm.
\begin{eqnarray}
  \left \| \frac{\mathrm{e}^{-\Delta\tau \hat{h}[m]} |\Psi_n\rangle}
 {\sqrt{\langle \Psi_n|\mathrm{e}^{-2\Delta\tau \hat{h}[m]}|\Psi_n\rangle}} 
  - \mathrm{e}^{-i\Delta\tau \hat{A}_n[m]} |\Psi_n\rangle \right\|^2 \,\, . 
\label{min}
\end{eqnarray}

\subsection*{Nonlocal condition for imaginary-time evolution operators}
We express the Hermitian operator $\hat{A}_n[m]$ as a linear combination of
the $D$-th order tensor products of Pauli operators $\{\hat{I}_l, \hat{\sigma}_{X,l}, \hat{\sigma}_{Y,l}, \hat{\sigma}_{Z,l}\}$ acting on the $l$-th qubit as 
\begin{eqnarray}
\hat{A}_n[m]
={\sum_{l_{k+1}, \cdots , l_D\in \mathbb{L}_m}}^{\prime} \sum_{i_1\cdots i_D} &&
a^{(n)}_{i_1\cdots i_D, l_{1} \cdots l_D } [m] \nonumber  \\
&&\hat{\sigma}_{i_1, l_1(m)} \otimes \cdots
\otimes \hat{\sigma}_{i_D,l_D},
\label{eq:ela}
\end{eqnarray}
where the prime on the first summation symbol indicates removing the double counting of the repeated tensors. We defined $\mathbb{L}_m$ as the set of $N_{\mathbb{L}_m}$ qubits, each of which interact with those in the partial Hamiltonian $\hat{h}[m]$; however, it is not contained in $\hat{h}[m]$.
The parameter $D$, which is called the domain size, satisfies $k \leqq D \leqq k + N_{\mathbb{L}_m}$, where we assumed the partial Hamiltonian $\hat{h}[m]$ to be written by a tensor product of the $k$-th order. $\{l_1(m),\cdots,l_{k}(m)\}$ is the set of qubits contained in the partial Hamiltonian $\hat{h}[m]$. The summation in Eq. (\ref{eq:ela}) is taken over all combinations of $D - k$ qubits, $\{l_{k+1}(m),\cdots,l_{D}(m)\}$, and chosen from $\mathbb{L}_m$. $D$ is an input parameter that represents the level of approximation; a larger $D$ indicates that the imaginary-time evolution operator is expressed using higher-order tensor products and the residual norm in Eq. (\ref{min}) shows a smaller value, which leads to a better approximation. 
We consider a scenario where the domain size $D$ incorporates all elements in $\mathbb{L}_m$, namely $D=k+N_{\mathbb{L}_m}$, and then Eq. (\ref{eq:ela}) reproduces the operator $A_n [m]$ introduced in Ref. \cite{motta2020determining}. This implies that Eq. (\ref{eq:ela})  is a natural extension of the approximation introduced in Ref. \cite{motta2020determining}. We call the method for determining the operator $A_n [m]$ defined in Ref. \cite{motta2020determining} local approximation (LA) for comparison with later approximation. Then, we refer to the method defined in Eq. (\ref{eq:ela}) as extended local-approximation (eLA).
The following notation is used to indicate the domain size $D$: e.g., LA with $D = 6$ is denoted by LA-D6. Note that, for LA, it is a well-defined approximation only when the domain size $D=k+N_{\mathbb{L}_m}$, and the value of $D$ that can be taken is limited by the Hamiltonian. In addition, note that eLA can remove such constraints on the Hamiltonian and flexibly determine the parameter $D$ by considering the linear combination for qubits. This flexibility is obvious in the max-cut problem of the fully connected graph. Solving the minimization problem in Eq. (\ref{min}) to determine the coefficients $a^{(n)}_{\{i,l\}}[m]$ results in the linear equation $S^{(n)}a^{(n)}[m]=b^{(n)}[m]$, which can be solved using a classical computer. 
Here, $S^{(n)}_{\{i,l_i\}\{j,l_j\}}=\langle \Psi_n |\hat{\sigma}^{\dagger}_{\{i,l_i\}} \hat{\sigma}_{\{j,l_j\}} |\Psi_n\rangle$ and $b^{(n)}_{\{i,l_i\}}[m]=\langle \Psi_n |\hat{\sigma}^{\dagger}_{\{i,l_i\}} \hat{h}[m] |\Psi_n\rangle$. Figure \ref{img:circ_depth}(a) shows a schematic of the quantum circuit representing one imaginary-time step of LA. In LA, the operator of the imaginary-time evolution is approximated by the tensor products of Pauli operators up to the $D$-th order; therefore, $4^D$ gate operations are required for each partial Hamiltonian. The total number of gate operations for one step of the imaginary-time evolution is $N_{\rm ham}4^D$. Table \ref{tab:circ_depth} summarizes the size of the linear equation of the LA per step of the imaginary-time evolution and the number of gate operations per qubit, where $N_{\rm bit}$ is the total number of qubits.
\begin{table}[htb]
  \begin{center}
    \caption{Scaling of the size of the matrix $S^{(n)}$ and the number of gate operations per qubit of the linear equation of LA and NLA per imaginary-time step}
    \begin{tabular}{lcc} \hline
      Method\qquad & \begin{tabular}{l}Scaling of the size of \\the linear equation\end{tabular}\qquad& 
       \begin{tabular}{l}Scaling of the gate \\operations per qubit \end{tabular} \rule[0mm]{0mm}{3mm} \\ \hline 
      LA & $4^D$ & $4^{D}N_{\rm ham} D/N_{\rm bit} $ \rule[0mm]{0mm}{5mm} \\
      NLA & $4^{D} {}_{N_{\rm bit}} C _{D}$ & $4^{D} {}_{N_{\rm bit}-1} C _{D-1}$  \rule[0mm]{0mm}{5mm} \\ \hline
    \end{tabular}
    \label{tab:circ_depth}
  \end{center}
\end{table}

Furthermore, this study proposes another approximation method for $\hat{A}_n$ in the following form: 
\begin{eqnarray}
 \hat{A}_n=
{\sum_{l_1, \cdots l_D}}^{\prime} \sum_{i_1, \cdots i_D} a^{(n)}_{i_1\cdots i_D, l_1 \cdots l_D} \hat{\sigma}_{i_1, l_1} 
 \otimes \cdots \otimes \hat{\sigma}_{i_D,l_D}.
 \label{eq:nla}
\end{eqnarray}
The difference from Eq. (\ref{eq:ela}) is that we remove the restriction on the set $\{l_1(m),\cdots l_{k}(m)\}$ and extend the summation over qubits to incorporate all possible combinations of $D$ qubits $\{l_1(m),\cdots l_{D}(m)\}$. We call this an NLA. As per this definition, we expand the Hermitian operator, $\hat{A}_n$, using tensor products of Pauli operators over all qubit combinations. In LA and eLA, the tensor product space describing $\hat{A}_n[m]$ is different depending on $m$, which is the partial Hamiltonian. The NLA has a notable feature in that the tensor product space that describes $\hat{A}[m]$ is the same for all $m$. Table  \ref{tab:circ_depth} lists the size of the linear equations of the NLA per step of the imaginary-time evolution and the number of gate operations per qubit, where the NLA requires only $4^D$ unitary operators in ${}_{N_{\rm bit}} C _{D}$ combinations for the quantum circuit in the first step of the imaginary-time evolution. Figure \ref{img:circ_depth}(b) shows the schematic of the quantum circuit of the NLA for one step of the imaginary-time evolution (for $D = 2$).

\section*{Results}
To clarify the accuracy and effectiveness of NLA, we applied it to the max-cut problem, which is an NP-hard problem. The Hamiltonian of the max-cut problem in qubit representation is given in the following form containing second-order tensor products \cite{Lucas2014IsingFO}.
\begin{eqnarray*}
\hat{H} = - \sum_{(i,j)\in E} d_{i,j}\frac{1-\hat{\sigma}_{Z,i}\hat{\sigma}_{Z,j}}{2}
\end{eqnarray*}
As for the max-cut problem, we considered typical graphs such as 3-regular and fully connected graphs. The 3-regular graphs have three connected edges at every vertex, where $E$ is the set of edges contained in the graph and $d_{i,j}$ is the weight of the edges connecting the $i$th and $j$th vertices.

\subsection*{Reduction effect of circuit depth}
The circuit depths when LA and NLA are applied to the max-cut problem are shown in Fig. \ref{img:circ_depth}(c) for the 3-regular graph and Fig. \ref{img:circ_depth}(d) for the fully-connected graph because different graphs of the max-cut problem change the number of the partial Hamiltonian $N_{\rm ham}$; the necessary circuit depths for each approximation change correspondingly. In Fig. \ref{img:circ_depth}(c) and (d), the circuit depth calculated using Qiskit\cite{aleksandrowicz2019qiskit} is plotted with points, and the plotted points are extrapolated. In the case of $k$-regular graphs, the number of the partial Hamiltonians is given by $N_{\rm ham}=kN_{\rm bit}/2$. It increases linearly with the number of vertices $N_{\rm bit}$ so that the number of gate operations per qubit does not depend on the number of qubits, as listed in Table \ref{tab:circ_depth}. Thus, we extrapolated using $y={\rm const.}$. In NLA, regardless of the structure of the Hamiltonian, the number of gate operations per qubit is scaled by $\mathcal{O}({N_{\rm bit}}^{D-1})$ with respect to the number of qubits $N_{\rm bit}$ because all combinations of ${}_{N_{\rm bit}} C _{D}$ are taken for gate operations including the $D$-th order tensor product. In Figure \ref{img:circ_depth}(c), the circuit depth of the NLA is extrapolated by the function fitted by $f(x)=x^{D-1}$.

Note that in LA, $D = 3$, 4, and 5 are not well defined in the 3-regular graph. Thus, $D = 6$ is required, and $4^6 = 4096$ gate operations are necessary for the imaginary-time evolution of one partial Hamiltonian, which leads to a deeper circuit depth and difficulty in implementation on NISQ devices. In addition, the circuit depth required for LA-D6, compared to NLA-D2, NLA-D3, etc., is considerably higher in the region with a small number of qubits. The circuit depth of the NLA becomes deeper than that of LA in the region where the number of qubits increases.

In Figure \ref{img:circ_depth}(d), LA-D2 and eLA-D3 are not shown for the fully connected graph ($N_{\rm ham}= {}_{N_{\rm bit}} C _{2}$) because the circuit depth of LA-D2 is equal to that of the NLA-D2, and that of eLA-D3 is equal to that of NLA-D3. In addition, because the domain size has to be $D=N_{\rm bit}$ in LA, which is the exact imaginary-time evolution in a fully connected graph, and the circuit depth increases exponentially with respect to the number of qubits. In NLA, it can be scaled down to the linear or quadratic function with respect to the number of qubits. This result indicates that the NLA and eLA are efficient in reducing the circuit depth, especially when the number of partial Hamiltonians increases; further, these algorithms are effective for NISQ devices.

\begin{figure*}[ht]
   \begin{center}
       \includegraphics[bb=0 0 1568 950,scale=0.33]{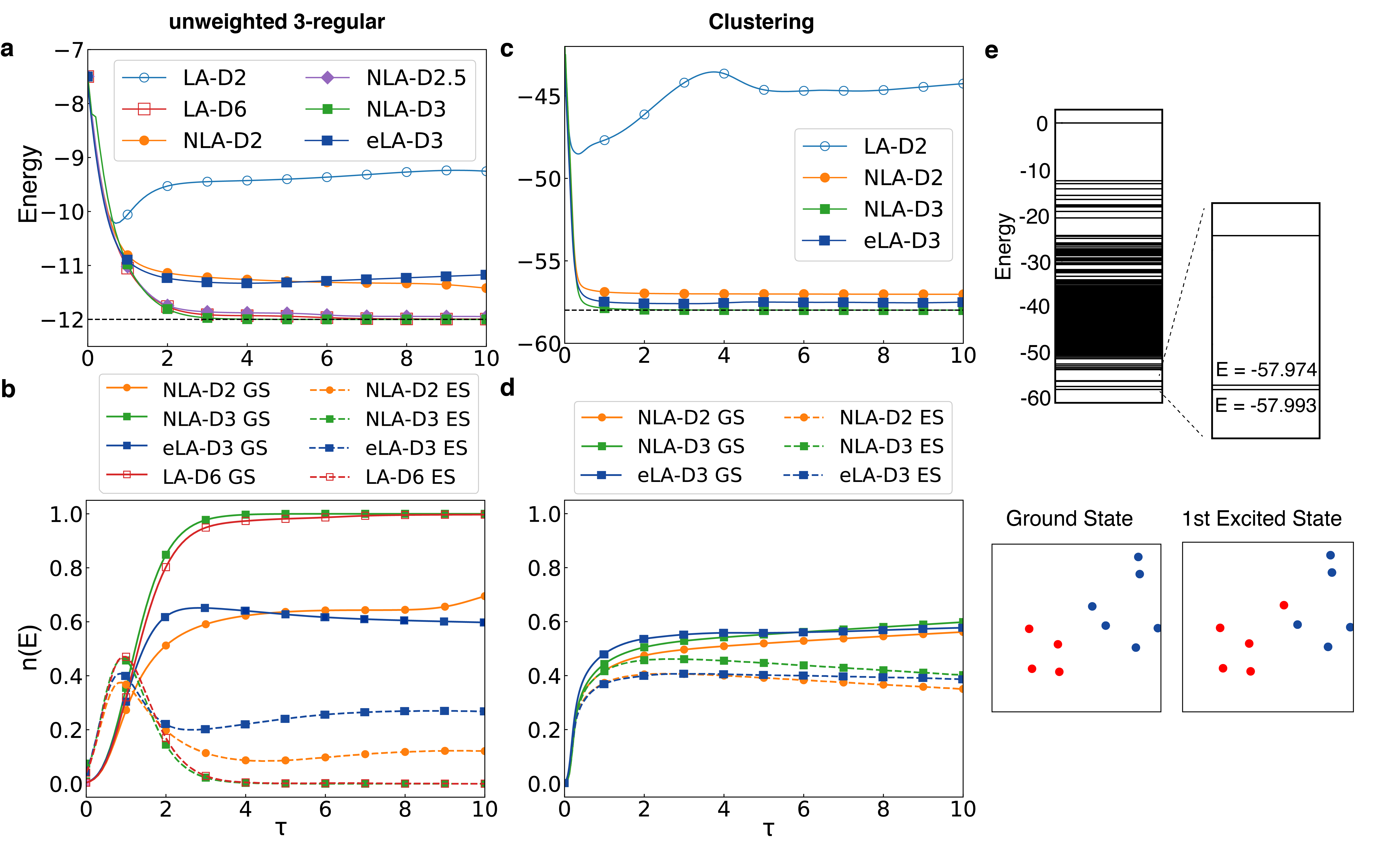}
       \caption{Energy $E$(a) and component proportions of the state $n(E)$(b) in the QITE method to the max-cut problem for an unweighted 3-regular graph with ten vertices. The ground state is denoted by GS and the first excited state by ES. The energy $E$(c) and component proportions $n(E)$(d) of the QITE method for a weighted fully connected graph with ten vertices. (e) The total energy level diagram of the weighted fully connected graph and the eigenstates corresponding to the ground state and the first excited state (divided into two regions, red and blue).
       \label{img:cmp_prev_rslt}}
   \end{center}
\end{figure*}

\subsection*{Calculation accuracy}
Simulations were performed after modifying the code provided in Ref. \cite{motta2020determining}. As an initial state, we adopted a state in which all states were superimposed with equal {\it a priori} weights. We adopt a figure of merit to discuss the accuracy of the QITE method. 
\begin{eqnarray*}
r = \lim_{\tau \to \infty}
\frac{\langle \Psi(\tau)|\hat{H}|\Psi(\tau)\rangle}{E_{\rm GS}}.
\end{eqnarray*}

The first target of the max-cut problem is an unweighted 3-regular graph with ten vertices, where $E_{\rm GS}$ is the energy of the ground state, and it is obtained from the exact diagonalization. The energy of the ground state is $E_{\rm GS}=-12$. 
It is known that designing a classical algorithm that achieves $r > 331/332$ for an unweighted 3-regular graph is an NP-hard problem \cite{berman1999on}. Further, the approximation accuracy of the current classical algorithm is $r\approx 0.9326$ \cite{goemans1995improved}. Figure \ref{img:cmp_prev_rslt} (a) shows the imaginary-time dependence of the energy. The imaginary-time step was set to $\Delta\tau = 0.01$. In LA-D2, as the imaginary-time $\tau$ increased, the energy decreased exponentially in the beginning and converged to around $-9$, which is higher than the exact solution by about $3$. Another important point is that the energy does not monotonically decreases along the imaginary-time evolution. This behavior indicates that the conversion of the operator of the imaginary-time evolution to the unitary operators is less accurate in expanding it in the space of  LA-D2. Furthermore, the LA-D6 calculation result shows $E = -11.99$, which is the energy almost equal to the exact solution. We found that an approximation accuracy in the eLA-D3 is $E = -11.17$ ($r = 0.93$) (the lowest value is $E = -11.33$ ($r = 0.94$)); in NLA-D2, $E = -11.42$ ($r = 0.95$); and in NLA-D3, $E = -12.00$ ($r = 1.00$). We found that eLA-D3 had an approximation accuracy similar to that of the classical algorithm, and NLA-D2 had already exceeded the approximation accuracy of the classical algorithm. 
Note that eLA and NLA monotonically decrease the energy along the imaginary-time evolution with sufficiently good accuracy compared to LA-D2. This behavior was confirmed not only for NLA-D2 but also for NLA-D3 and others. As can be seen from Fig. \ref{img:circ_depth} (c), in LA-D6, the circuit depth of one imaginary-time step is 369757, while the circuit depth in the NLA-D2 is 789. This implies the circuit depth of NLA can be significantly shallower than that of LA.
 
 While NLA-D3 has extremely high accuracy, its circuit depth increases with a quadratic function with respect to the number of qubits. Then, we developed NLA-D2.5 to keep the scaling of the circuit depth as linear as NLA-D2 while maintaining the accuracy of NLA-D3, which is an approximation to expand the space of $\hat{A}_n$ to the space involving the second-order tensor products incorporated by NLA-D2 and the third-order tensor products by eLA-D3. Thus, by incorporating some portions of bases of eLA-D3 into those of NLA-D2, computational scaling can be made linear with respect to the number of qubits, which makes it applicable even in regions with a large number of qubits. 
Fig. \ref{img:circ_depth}(c) shows that the circuit depth is almost the same as that of NLA-D2 for 50 qubits or more, which means that the circuit depth can be significantly reduced compared to that of NLA-D3. In addition, the calculation result of NLA-D2.5 is $E = -11.95$ and $r = 0.99$, which gives a good approximation accuracy with a small circuit depth. 
 
Here, for further consideration, we decomposed the state $|\Psi(\tau)\rangle = \mathrm{e}^{-\tau\hat{H}}|\Psi(\tau = 0)\rangle$ into the eigenstate components of the Hamiltonian, and the calculated $n(E) \equiv \sum_{i} |\langle i|\Psi(\tau)\rangle|^2\delta(E-E_i)$ as a function of energy $E$ at each imaginary-time step $\tau$ is plotted in Fig. \ref{img:cmp_prev_rslt}(b) where $|i\rangle$ is the eigenstate of $\hat{H}$ and $E_i$ is the eigen-energy of $|i\rangle$. The ground state can be observed with probabilities of $n(E_{\rm GS}) = 0.60$ for eLA-D3 (at maximum, $n(E_{\rm GS}, \tau = 2.87) = 0.65)$, $n(E_{\rm GS}) = 0.69$ for NLA- D2, $n(E_{\rm GS}) = 0.97$ for NLA-D2.5, and $n(E_{\rm GS}) = 1.00$ for NLA-D3. The imaginary-time dependence of the probability of the first excited state is also plotted. For the first excited state, it is observed that the probability is amplified up to $\tau = 1$, and it starts to decrease, which increases the ground state probability.

Next, we deal with another computational model called a weighted fully connected graph (classification problem). The coupling constants $d_{i,j}$ were given by random numbers. The ground state energy is $E_{\rm GS} = -57.993$. In addition, the imaginary-time step is set to $\Delta\tau = 0.01$. In the classification problem, as shown in Fig. \ref{img:cmp_prev_rslt}(e), each graph vertex is colored red or blue. In LA-D2, as in the 3-regular graph, we observed that the energy does not necessarily decrease monotonically. The energy of eLA-D3 is lower than that of NLA-D2;  $E = -57.504$ ($r = 0.99$) for eLA-D3, 
$E = -57.026$($r = 0.98$) for NLA-D2, and $E = -57.985$($r = 0.99$) for NLA-D3 (Figure \ref{img:cmp_prev_rslt} (c)). 
From the viewpoint of the component analyses of the states, the ground state and the first excited state are pseudo-degenerate (Fig. \ref{img:cmp_prev_rslt}.(e)), and therefore, the probability of the first excited state remains at the same level as the ground state even around $\tau = 2$ when the energy converges sufficiently (Fig. \ref{img:cmp_prev_rslt}.(d)). In NLA, the first excited state gradually decays along with the imaginary-time evolution; however, a sufficiently long imaginary-time evolution is necessary. In particular, NLA-D2 behaves similarly to NLA-D3, and NLA-D2 is sufficiently accurate to obtain the ground state in actual applications.

\subsection*{Compression of imaginary-time steps}
\begin{figure*}[ht]
   \begin{center}
       \includegraphics[bb=0 0 1140 873,scale=0.4]{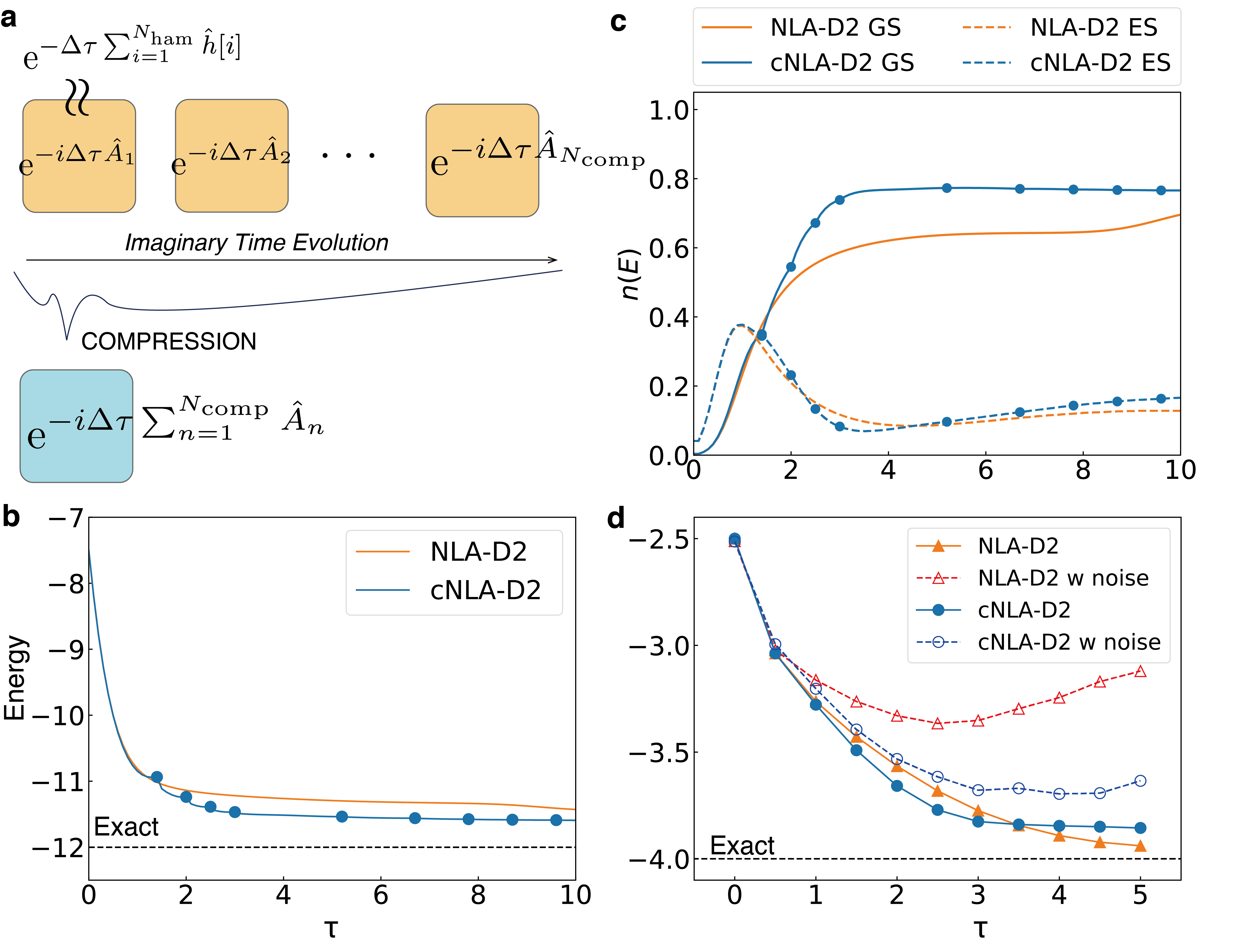}
       \caption{(a) Schematic of the compression of the imaginary-time step. energy $E$(b) and component of the eigenstate $n(E)$(c) in the imaginary-time evolution with and without compression of the imaginary-time step in the max-cut problem of an unweighted 3-regular graph with ten vertices. The compressed point $N_{{\rm comp}}$ is plotted with circles. (d) Results of the simulation with noise for the max-cut problem in an unweighted graph with four vertices.
       \label{img:nla_cluster}}
   \end{center}
\end{figure*}
The approximation accuracy of the NLA and its circuit depth have been discussed. The ``compression of imaginary-time steps'' is introduced in this section for further reduction of the number of gate operations in NLA. Figure \ref{img:nla_cluster} (a) shows a schematic of the compression technique. When the imaginary-time step $\Delta\tau$ is sufficiently small, the time-evolution operators can be compressed into a single exponential form via the reverse Suzuki--Trotter decomposition
\begin{eqnarray*}
\prod_{n = 1}^{N_{\rm comp}} \exp(-i\Delta\tau \hat{A_n})
= \exp(-i \Delta \tau \sum_{n = 1}^{N_{\rm comp}} \hat{A_n})
+ \mathcal{O}(\Delta\tau^2),
\end{eqnarray*}
where $N_{\rm comp}$ is the number of compressed steps. It is necessary to choose an appropriate $N_{\rm comp}$ within the range that guarantees sufficient accuracy for the Suzuki--Trotter decomposition because its accuracy decreases if the $N_{\rm comp}$ becomes large. To determine the specific $N_{\rm comp}$ in this work, we increased the $N_{\rm comp}$ parameter by one at every time-evolution step until the total energy increases. In actual QITE calculations, $N_{\rm comp}$ is not necessarily a constant throughout the calculation. This method enables the reduction of quantum circuits to be as small as $1/N_{\rm comp}$.

The graph used for the calculation is the same as that in Figures \ref{img:cmp_prev_rslt}(a) and (b), which is a 3-regular graph with ten vertices. Figure \ref{img:nla_cluster} shows the results of the compression technique for the QITE. In Figure \ref{img:nla_cluster}(b), the time the compression ended is plotted as a blue circle. In the case of Fig. \ref{img:nla_cluster}(b), the quantum circuit depth is significantly reduced by the compression technique to four compressed imaginary-time steps, and the energy at $\tau = 10$ is $E = -11.43$ ($r = 0.95$) without and $E = -11.59$ ($r = 0.97$) with the compression technique. We found that sufficient accuracy was achieved regardless of the compression, which indicates compression does not affect the results. It may be assumed that the compressed technique has a lower energy than that of the uncompressed calculation; a detailed investigation revealed that this was attributed to the accidental acceleration of the convergence by compression. 
Figure \ref{img:nla_cluster} (c) plots the component analyses of the wavefunctions during the imaginary-time evolution with and without the compression method. Finally, the probability of obtaining a ground state is $n(E_{\rm min}) = 0.76$  with and $n(E_{\rm min}) = 0.73$ without the compression technique.

The ``compression of imaginary-time steps'' is effective in reducing the circuit depth, and simultaneously, it reduces the noise associated with the gate operations. We discuss the results of the simulation with noise.
The actual qubits are currently connected only with neighboring sites; however, in this study, we simulated a fully connected model. For implementation on an actual quantum computer, in which only adjacent sites are connected, a \textsf{SWAP} gate can be used with an overhead of $\mathcal{O}(\sqrt{N_{\rm bit}})$\cite{cheung2007translation}. For example, QAOA uses a \textsf{SWAP} network \cite{kivlichan2018quantum,babbush2018low} to implement a $\mathcal{O}(N_{\rm bit})$ overhead \cite{crooks2018performance}. The error model of the gate was constructed from the thermal relaxation time $(T_1, T_2)=(100{\rm \mu s}, 80{\rm \mu s})$, and the gate time $(T_{g1}, T_{g2})=(0.02{\rm ns}, 0.1{\rm ns})$. The noise simulation was performed by introducing the readout errors $(p_{00}, p_{01}, p_{10}, p_{11})=(0.995,0.005,0.02,0.98)$. These parameters were assumed to be close to the actual values of IBMQ\cite{ibmq}.
Figure \ref{img:nla_cluster} (d) shows the simulation results of the max-cut problem for an unweighted graph with four vertices. The coefficients $a^{(n)}_{\{i, l\}}$ in Eq. (\ref{eq:nla}) for the noisy calculation are the same as those for the non-noisy calculation. The noiseless condition without compression results in $E = -3.94$, which is close to the exact solution $E = -4.00$ around $\tau = 5$. However, the circuit depth is 922 ($\Delta\tau = 0.5$), and the simulation result with noise is $E = -3.13$, which is far from the exact solution. This gap was attributed to the accumulation of errors caused by an increase in circuit depth. The result with compression is $E = -3.85$ in the case without noise; however, the circuit depth is 163, and the effect of noise is expected to be less sensitive. In fact, the simulation result with noise is $E = -3.63$, which shows that the noise can be reduced with compression. Thus, it has been shown that the ``compression'' method of quantum circuits has an advantage of reducing the accumulation of errors.

\section*{Concluding Remarks}
In this study, we proposed two-step approximation methods based on nonlocality: eLA and NLA. We applied them to the Max-cut problem of an unweighted 3-regular graph and a weighted fully-connected graph, and comparatively validated the performances of LA, eLA, and NLA. 
We found that NLA requires significantly less circuit depth than LA while maintaining the same level of computational accuracy. For example, when we request the classical approximation limit in the QITE calculations, the circuit depth required for a single imaginary-time step can be significantly reduced from 369757 for LA to 789 for NLA when applying it to a 3-regular graph, and from about 314000 for LA to 789 for NLA when applying it to a fully connected graph. 
Further, we developed a ``compression'' technique of the imaginary-time evolution steps to further reduce the circuit depth in the QITE method. With this compression method, we succeeded in further reducing the circuit depth. We showed that the reduction in circuit depth using this compression method has a secondary effect of reducing the accumulation of error caused by the gate operation. Thus, it is an effective method for realization on NISQ devices. 
The eLA, NLA, and compression methods introduced in this study enable us to reduce the circuit depth and the accumulation of error caused by the gate operation significantly and have paved the way for the realization of the QITE method on NISQ devices.

\section*{Data availability}
The datasets generated during and/or analysed during the current study are
available from the corresponding author on reasonable request.

\bibliographystyle{junsrt}
\bibliography{paper.bib}

\begin{thebibliography}{10}

\bibitem{feynman1999simulating}
Richard~P Feynman.
\newblock Simulating physics with computers.
\newblock {\em Int. J. Theor. Phys}, Vol.~21, No. 6/7, 1999.

\bibitem{deutsch1992rapid}
David Deutsch and Richard Jozsa.
\newblock Rapid solution of problems by quantum computation.
\newblock {\em Proceedings of the Royal Society of London. Series A:
  Mathematical and Physical Sciences}, Vol. 439, No. 1907, pp. 553--558, 1992.

\bibitem{grover1996fast}
Lov~K Grover.
\newblock A fast quantum mechanical algorithm for database search.
\newblock In {\em Proceedings of the twenty-eighth annual ACM symposium on
  Theory of computing}, pp. 212--219, 1996.

\bibitem{shor1999polynomial}
Peter~W Shor.
\newblock Polynomial-time algorithms for prime factorization and discrete
  logarithms on a quantum computer.
\newblock {\em SIAM review}, Vol.~41, No.~2, pp. 303--332, 1999.

\bibitem{arute2019quantum}
Frank Arute, Kunal Arya, Ryan Babbush, Dave Bacon, Joseph~C Bardin, Rami
  Barends, Rupak Biswas, Sergio Boixo, Fernando~GSL Brandao, David~A Buell,
  et~al.
\newblock Quantum supremacy using a programmable superconducting processor.
\newblock {\em Nature}, Vol. 574, No. 7779, pp. 505--510, 2019.

\bibitem{hyafil1973graph}
Laurent Hyafil and Ronald~L Rivest.
\newblock {\em Graph partitioning and constructing optimal decision trees are
  polynomial complete problems}.
\newblock IRIA. Laboratoire de Recherche en Informatique et Automatique, 1973.

\bibitem{lawler1985traveling}
Eugene~L Lawler.
\newblock The traveling salesman problem: a guided tour of combinatorial
  optimization.
\newblock {\em Wiley-Interscience Series in Discrete Mathematics}, 1985.

\bibitem{ho2004genace}
Nhu~Binh Ho and Joc~Cing Tay.
\newblock Genace: an efficient cultural algorithm for solving the flexible
  job-shop problem.
\newblock In {\em Proceedings of the 2004 Congress on Evolutionary Computation
  (IEEE Cat. No. 04TH8753)}, Vol.~2, pp. 1759--1766. IEEE, 2004.

\bibitem{karp1972reducibility}
Richard~M Karp.
\newblock Reducibility among combinatorial problems.
\newblock In {\em Complexity of computer computations}, pp. 85--103. Springer,
  1972.

\bibitem{durr1996quantum}
Christoph Durr and Peter Hoyer.
\newblock A quantum algorithm for finding the minimum.
\newblock {\em arXiv preprint quant-ph/9607014}, 1996.

\bibitem{baritompa2005grover}
William~P Baritompa, David~W Bulger, and Graham~R Wood.
\newblock Grover's quantum algorithm applied to global optimization.
\newblock {\em SIAM Journal on Optimization}, Vol.~15, No.~4, pp. 1170--1184,
  2005.

\bibitem{preskill2018quantum}
John Preskill.
\newblock Quantum computing in the nisq era and beyond.
\newblock {\em Quantum}, Vol.~2, p.~79, 2018.

\bibitem{peruzzo2014variational}
Alberto Peruzzo, Jarrod McClean, Peter Shadbolt, Man-Hong Yung, Xiao-Qi Zhou,
  Peter~J Love, Al{\'a}n Aspuru-Guzik, Jeremy~LO’brien.
\newblock A variational eigenvalue solver on a photonic quantum processor.
\newblock {\em Nature communications}, Vol.~5, p. 4213, 2014.

\bibitem{mcclean2016theory}
Jarrod~R McClean, Jonathan Romero, Ryan Babbush, and Al{\'a}n Aspuru-Guzik.
\newblock The theory of variational hybrid quantum-classical algorithms.
\newblock {\em New Journal of Physics}, Vol.~18, No.~2, p. 023023, 2016.

\bibitem{farhi2014quantum}
Edward Farhi, Jeffrey Goldstone, and Sam Gutmann.
\newblock A quantum approximate optimization algorithm.
\newblock {\em arXiv preprint arXiv:1411.4028}, 2014.

\bibitem{otterbach2017unsupervised}
JS~Otterbach, R~Manenti, N~Alidoust, A~Bestwick, M~Block, B~Bloom, S~Caldwell,
  N~Didier, E~Schuyler Fried, S~Hong, et~al.
\newblock Unsupervised machine learning on a hybrid quantum computer.
\newblock {\em arXiv preprint arXiv:1712.05771}, 2017.

\bibitem{moll2018quantum}
Nikolaj Moll, Panagiotis Barkoutsos, Lev~S Bishop, Jerry~M Chow, Andrew Cross,
  Daniel~J Egger, Stefan Filipp, Andreas Fuhrer, Jay~M Gambetta, Marc Ganzhorn,
  et~al.
\newblock Quantum optimization using variational algorithms on near-term
  quantum devices.
\newblock {\em Quantum Science and Technology}, Vol.~3, No.~3, p. 030503, 2018.

\bibitem{wang2018quantum}
Zhihui Wang, Stuart Hadfield, Zhang Jiang, and Eleanor~G Rieffel.
\newblock Quantum approximate optimization algorithm for maxcut: A fermionic
  view.
\newblock {\em Physical Review A}, Vol.~97, No.~2, p. 022304, 2018.

\bibitem{Hadfield2019FromTQ}
Stuart Hadfield, Zhihui Wang, Bryan O'Gorman, Eleanor~G. Rieffel, Davide
  Venturelli, and Rupak Biswas.
\newblock From the quantum approximate optimization algorithm to a quantum
  alternating operator ansatz.
\newblock {\em Algorithms}, Vol.~12, p.~34, 2019.

\bibitem{guerreschi2019qaoa}
Gian~Giacomo Guerreschi and Anne~Y Matsuura.
\newblock Qaoa for max-cut requires hundreds of qubits for quantum speed-up.
\newblock {\em Scientific reports}, Vol.~9, , 2019.

\bibitem{mcclean2018barren}
Jarrod~R McClean, Sergio Boixo, Vadim~N Smelyanskiy, Ryan Babbush, and Hartmut
  Neven.
\newblock Barren plateaus in quantum neural network training landscapes.
\newblock {\em Nature communications}, Vol.~9, No.~1, pp. 1--6, 2018.

\bibitem{Lucas2014IsingFO}
Andrew Lucas.
\newblock Ising formulations of many np problems.
\newblock {\em ArXiv}, Vol. abs/1302.5843, , 2014.

\bibitem{mcardle2019variational}
Sam McArdle, Tyson Jones, Suguru Endo, Ying Li, Simon~C Benjamin, and Xiao
  Yuan.
\newblock Variational ansatz-based quantum simulation of imaginary time
  evolution.
\newblock {\em npj Quantum Information}, Vol.~5, No.~1, pp. 1--6, 2019.

\bibitem{stokes2019quantum}
James Stokes, Josh Izaac, Nathan Killoran, and Giuseppe Carleo.
\newblock Quantum natural gradient.
\newblock {\em arXiv preprint arXiv:1909.02108}, 2019.

\bibitem{david2020avoiding}
Wierichs David, Gogolin Christian, and Kastoryano Michael.
\newblock Avoiding local minima in variational quantum eigensolvers with the
  natural gradient optimizert.
\newblock {\em arXiv preprint arXiv:2004.14666}, 2020.

\bibitem{motta2020determining}
Mario Motta, Chong Sun, Adrian~TK Tan, Matthew~JO’Rourke, Erika Ye, Austin~J
  Minnich, Fernando~GSL Brand{\~a}o, Garnet Kin-Lic Chan.
\newblock Determining eigenstates and thermal states on a quantum computer
  using quantum imaginary time evolution.
\newblock {\em Nature Physics}, Vol.~16, No.~2, pp. 205--210, 2020.

\bibitem{yeter2019practical}
K{\"u}bra Yeter-Aydeniz, Raphael~C Pooser, and George Siopsis.
\newblock Practical quantum computation of chemical and nuclear energy levels
  using quantum imaginary time evolution and lanczos algorithms.
\newblock {\em arXiv preprint arXiv:1912.06226}, 2019.

\bibitem{beach2019making}
Matthew~JS Beach, Roger~G Melko, Tarun Grover, and Timothy~H Hsieh.
\newblock Making trotters sprint: A variational imaginary time ansatz for
  quantum many-body systems.
\newblock {\em Physical Review B}, Vol. 100, No.~9, p. 094434, 2019.

\bibitem{jain1999data}
Anil~K Jain, M~Narasimha Murty, and Patrick~J Flynn.
\newblock Data clustering: a review.
\newblock {\em ACM computing surveys (CSUR)}, Vol.~31, No.~3, pp. 264--323,
  1999.

\bibitem{jain1988algorithms}
Anil~K Jain and Richard~C Dubes.
\newblock {\em Algorithms for clustering data}.
\newblock Prentice-Hall, Inc., 1988.

\bibitem{bravyi2002fermionic}
Sergey~B. {Bravyi} and Alexei~Yu. {Kitaev}.
\newblock Fermionic quantum computation.
\newblock {\em Annals of Physics}, Vol. 298, No.~1, pp. 210--226, 2002.

\bibitem{jordan1928uber}
P.~{Jordan} and E.~P. {Wigner}.
\newblock ^^c3^^9cber das paulische ^^c3^^84quivalenzverbot.
\newblock {\em European Physical Journal}, Vol.~47, No.~9, pp. 631--651, 1928.

\bibitem{aleksandrowicz2019qiskit}
Gadi Aleksandrowicz, Thomas Alexander, Panagiotis Barkoutsos, Luciano Bello,
  Yael Ben-Haim, D~Bucher, FJ~Cabrera-Hern{\'a}ndez, J~Carballo-Franquis,
  A~Chen, CF~Chen, et~al.
\newblock Qiskit: An open-source framework for quantum computing.
\newblock {\em Accessed on: Mar}, Vol.~16, , 2019.

\bibitem{berman1999on}
Piotr {Berman} and Marek {Karpinski}.
\newblock On some tighter inapproximability results (extended abstract).
\newblock In {\em ICAL '99 Proceedings of the 26th International Colloquium on
  Automata, Languages and Programming}, pp. 200--209, 1999.

\bibitem{goemans1995improved}
Michel~X. {Goemans} and David~P. {Williamson}.
\newblock Improved approximation algorithms for maximum cut and satisfiability
  problems using semidefinite programming.
\newblock {\em Journal of the ACM}, Vol.~42, No.~6, pp. 1115--1145, 1995.

\bibitem{cheung2007translation}
Donny Cheung, Dmitri Maslov, and Simone Severini.
\newblock Translation techniques between quantum circuit architectures.
\newblock In {\em Workshop on Quantum Information Processing}, 2007.

\bibitem{kivlichan2018quantum}
Ian~D Kivlichan, Jarrod McClean, Nathan Wiebe, Craig Gidney, Al{\'a}n
  Aspuru-Guzik, Garnet Kin-Lic Chan, and Ryan Babbush.
\newblock Quantum simulation of electronic structure with linear depth and
  connectivity.
\newblock {\em Physical review letters}, Vol. 120, No.~11, p. 110501, 2018.

\bibitem{babbush2018low}
Ryan Babbush, Nathan Wiebe, Jarrod McClean, James McClain, Hartmut Neven, and
  Garnet Kin-Lic Chan.
\newblock Low-depth quantum simulation of materials.
\newblock {\em Physical Review X}, Vol.~8, No.~1, p. 011044, 2018.

\bibitem{crooks2018performance}
Gavin~E. {Crooks}.
\newblock Performance of the quantum approximate optimization algorithm on the
  maximum cut problem.
\newblock {\em arXiv preprint arXiv:1811.08419}, 2018.

\bibitem{ibmq}
{IBM Quantum Experience Web Site}.
\newblock \url{. https://quantum-computing.ibm.com/}.

\end{thebibliography}

\section*{Acknowledgement}
This research was supported by MEXT as an Exploratory Challenge on Post-K computer (Frontiers of Basic Science: Challenging the Limits) and by Grants-in-Aid for Scientific Research (A) (Grant Numbers 18H03770) from JSPS (Japan Society for the Promotion of Science).

\section*{Author contributions}
H. N. and Y. M. conceived the general idea. H. N. modified the code provided in prior work. H.N. and T. K. developed the code for noisy simulation. Numerical simulations were performed by H.N. All authors contributed equally to the manuscript preparation and presentation of results.

\end{document}